\documentclass[pra,showpacs,preprintnumbers,amsmath,aps]{revtex4}
\usepackage{graphics}

\def\be{\begin{equation}}
\def\ee{\end{equation}}
\def\bea{\begin{eqnarray}}
\def\eea{\end{eqnarray}}
\def\bse{\begin{subequations}}
\def\ese{\end{subequations}}
\def\bma{\begin{mathletters}}
\def\ema{\end{mathletters}}
\def\C{\hbox{$\mit I$\kern-.6em$\mit C$}}
\def\identity{{\rlap{1} \hskip 1.6pt \hbox{1}}}

\begin{document}

\title{Giant Gravitons in type IIA PP-wave Background}
\author{Yi-Xin Chen}
\email{yxchen@zimp.zju.edu.cn}
\author{Jing Shao}
\email{jingshao@zju.edu.cn}
\affiliation{Zhejiang Institute of
Modern Physics and Department of Physics, Zhejiang University,
Hangzhou 310027, P. R. China}

\date{\today}

\begin{abstract}
We examine giant gravitons with a worldvolume magnetic flux $q$ in
type IIA pp-wave background and find that they can move away from
the origin along $x^4$ direction in target space satisfying
$Rx^4=-q$. This nontrivial relation can be regarded as a
complementary relation of the giant graviton on IIA pp-wave and is
shown to be connected to the spacetime uncertainty principle. The
giant graviton is also investigated in a system of N D0-branes as
a fuzzy sphere solution. It is observed that $x^4$ enters into the
fuzzy algebra as a deformation parameter. Such a background
dependent Myers effect guarantees that we again get the crucial
relation of our giant graviton. In the paper, we also find a BIon
configuration on the giant graviton in this background.
\end{abstract}


\maketitle

\section{Introduction}

\indent

Giant gravitons have attracted substantial attention since the
pioneering paper \cite{susskind} of McGreevy, Susskind and
Toumbas. They found that a moving particle in $AdS\times S$
background can blow up into a spherical brane (giant graviton)
with the increasing angular momentum, and the stringy exclusion
principle \cite{stringy} can be naturally understood as the fact
that no particle is bigger than the sphere that contains it. Later
in \cite{miaoli} this remarkable mechanism is shown to be another
beautiful manifestation of the spacetime uncertainty principle
\cite{uncertainty} in string theory as well as M-theory. Some
important aspects of giant gravitons have been studied in
\cite{myers2}-\cite{jy2}. On the other hand, pp-wave as the
Penrose limit \cite{penrose} of $AdS\times S$ geometry provides
nontrivial but viable backgrounds to test $AdS/CFT$ correspondence
and to investigate string theories and various brane
configurations \cite{maldacena}-\cite{taylor}. So it is
interesting to investigate giant gravitons in such backgrounds.

Actually there has been much progress along this direction. It was
found that the giant gravitons on pp-wave generally have two
different guises. If the giant graviton wraps the boosted circle
of $AdS$ background, the Penrose limit gives a rotating lightcone
brane. In string theory it can be analyzed using the lightcone
worldsheet theory. The other type is the static spherical brane,
if the giant graviton is not wrapping the boosted circle before we
take the limit. These giant gravitons can be examined in D-brane
effective action or matrix model. In \cite{takayanagi} these two
kinds of giant gravitons on type IIB pp-wave have been
investigated using the Born-Infeld action and the string
worldsheet theory. It was found that giant gravitons with magnetic
flux can grow up from multiple D-strings. But there is a suspicion
that these D-strings will eventually decay into a giant graviton.
In M-theory, the giant gravitons on pp-wave are studied in BMN
matrix model \cite{maldacena}. They appear as concentric fuzzy
spheres with radius proportional to the lightcone momentum.

In this paper we focus on some configurations of giant graviton in
a type IIA pp-wave background, which is the KK reduction of the
maximally supersymmetric pp-wave of M-theory. We expect some new
features to arise in this background due to the compactification.
As a matter of fact, we find that the giant graviton on this
pp-wave can has an extra parameter $x^4$ as the location in
transverse space. When a magnetic flux on worldvolume is turned
on, the giant graviton will move away from the origin along $x^4$
while still preserving all the supersymmetries. It is also found
that the radius and the location of this giant graviton are
subject to a constraint $Rx^4=-q$. As will be elaborated later,
this relation is quite nontrivial since it is the core relation to
understand the giant graviton and the spacetime uncertainty
principle. Furthermore, since the magnetic flux on giant graviton
induces a D0-brane charge, we consider our giant graviton to be
composed of multi D0-branes and can be analyzed in the non-abelian
Born-Infeld action. We find the fuzzy sphere configuration
perfectly match our giant graviton, and the coordinate $x^4$
appear in the fuzzy algebra as a deformation parameter. It is also
noticed that our giant graviton admit a stringy description in the
matrix string theory on pp-wave. On the other hand, recently it
was found in \cite{chen} that spikes can grow on a dielectric
brane when an electric flux on the worldvolume is turned on. But
the background considered there is not a real supergravity
solution. So we wonder whether such a configuration is possible in
pp-wave background. In this paper we do find such a 1/8 BPS state
exists as a BIon configuration on a giant graviton.

The organization of this paper is as follows. In Section 2 we
examine the giant graviton in type IIA pp-wave background using
the abelian Born-Infeld action. The BPS solution with a magnetic
flux is obtained. The relation to the giant graviton of M-theory
is addressed, and the connection to the spacetime uncertainty
relation is also elucidated. In Section 3 we consider multiple
D0-branes blowing up into fuzzy spheres on the pp-wave as a
microscopic description of our giant graviton. We will also have a
brief discussion on how the giant graviton arise in matrix string
theory. The last section is devoted to the giant graviton with
electric flux in worldvolume. The BPS equation is written down for
the supersymmetric configuration. Solutions to this equation are
discussed and explanations are given. In the appendix, we
summarize some useful facts and conventions about the pp-wave
background considered in this paper.

\section{Giant gravitons on IIA pp-wave}

\indent

In this part we consider the giant graviton in the following type
IIA pp-wave background
\begin{eqnarray}
&&ds^2=-2dx^{-}dx^{+}-(\frac{\mu^2}{9}\sum_{i=1}^{4}(x^{i})^2+\frac{\mu^2}{36}\sum_{i'=5}^{8}(x^{i'})^2)(dx^{+})^2+
\sum_{I=1}^{8}(dx^{I})^2 \nonumber\\
&&F_{+123}=\mu, \quad F_{+4}=-\frac{\mu}{3},
\end{eqnarray}
by analyzing a spherical D2-brane from the Born-Infeld action.
More details about the background can be found in the appendix.
The low energy effective action of a D2-brane in general type IIA
backgrounds includes the Born-Infeld action and the Chern-Simons
terms, which reads as
\begin{equation}\label{DBI}
S_2=-T_2\int d^3\sigma(e^{-\phi}\sqrt{-det(P[G+B]_{ab}+\lambda
F_{ab})})+ \mu_2\int P[\sum{ C^{(n)}} e^{B} ]e^{\lambda F},
\end{equation}
where $T_2=2\pi/(2\pi l_s)^3 g_s$ is the D2-brane tension and
$\lambda=2\pi l^2_s$. As usual $P[\cdots]$ is used to clarify the
the pullbacks, and the potential $C^{(n)}$ of the RR field is
defined by $F^{(n+1)} =dC^{(n)}$. $\mu_2$ is the RR charge of the
D2-brane and supersymmetry requires that $\mu_2=\pm T_2$
corresponding to branes or antibranes. We should choose an
anti-brane if we assume $\mu$ to be positive in the following. For
the type IIA pp-wave background of our interest the action is
reduced to
\begin{equation}
S_2=-T_2\int d^3\sigma(\sqrt{-det(P[G]_{ab}+\lambda F_{ab})}+
\lambda C_0 F_{12}+ C_{012})
\end{equation}
The spherical D2-brane considered here has an embedding as
\begin{equation}
  x^{+} = t, \quad x^1 = R \sin\theta\cos\phi, \quad x^2 = R \sin\theta\sin\phi, \quad
  x^3 = R \cos\theta,
\end{equation}
where $\{t,\theta,\phi\}$ are chosen as the worldvolume
coordinates. We also switch on a worldvolume magnetic flux
\begin{equation}
F_{\theta \phi}=\lambda^{-1} q\sin\theta.
\end{equation}
Quantization of the magnetic flux requires
\begin{equation}
N=\frac{1}{2\pi}\int d\theta d\phi F_{\theta \phi}=2\lambda^{-1}q,
\end{equation}
where N is an integer and is interpreted as the number of
D0-branes bound to the D2-brane as can be seen from the coupling
of RR 1-form potential. We also have noticed that the transverse
scaler $x^4$ should be nonzero to make the embedding consistent
when we have a magnetic flux. Other scalers should be set to zero
since the gravity of the background provides a confining
potential. In our ansatz the RR potential can be chosen as
\begin{equation}
C_t=\frac{\mu}{3}x^4, \quad
C_{t\theta\phi}=-\frac{\mu}{3}R^3\sin\theta,
\end{equation}
and one can easily obtain the explicit form of the Lagrangian
density
\begin{equation}\label{abelian}
{\cal L}=-\frac{\mu}{3} T_2(\sqrt{(R^2+(x^4)^2)(R^4+q^2)}+x^4
q-R^3)\sin \theta.
\end{equation}
Varying $x^4$ and $R$ gives the equations of motion
\begin{eqnarray}\label{sph_equ1}
x^4\sqrt{R^4+q^2\over R^2+(x^4)^2}+q&=&0, \nonumber \\
\frac{2R}{3}\sqrt{R^2+(x^4)^2\over
R^4+q^2}+\frac{1}{3R}\sqrt{R^4+q^2\over R^2+(x^4)^2}-1&=&0.
\end{eqnarray}
It can be easily checked that these equations are the same as
those derived from the full field equations of motion
\cite{taylor} in our ansatz. So our embedding is consistent. The
equation (\ref{sph_equ1}) can be solved to give
\begin{equation}\label{ryconstraint}
Rx^4=-q.
\end{equation}
Substituting it back into the Hamiltonian ${\cal H}=-{\cal L}$, we
find that the energy of the configuration exactly equals to zero.
Actually this is just the giant graviton solution on IIA pp-wave.

It is interesting to notice that the size of the giant graviton is
inversely proportional to its location in $x^4$. This relation
will be shown to be connected to the spacetime uncertainty
principle. For now we can gain some insights by simple analysis.
First if $x^4$ is fixed, then $R$ is proportional to $q$. In other
words we can increase the radius of the giant graviton by adding a
magnetic flux to it. This is equivalent to adding D0-branes. Since
D0-branes repel each other, the increase of the radius is natural.
Second if $q$ is fixed, then $R$ is inversely proportional to
$x^4$. From the metric of pp-wave we can see that the gravity of
the background tends to make a spherical brane collapse. When
$x^4$ increases, this effect becomes stronger and the radius of
the giant graviton shrinks. We also notice that the giant graviton
with the same sign of $q$ only appear on one side of origin along
$x^4$. This asymmetry is due to the presence of KK gauge field of
the background.

Below we will show explicitly that the solution we obtained above
is actually a BPS object using the kappa symmetry projection. For
our brane embeding the kappa symmetry projection \cite{aganagic,
townsend} is
\begin{equation}\label{gamma}
\Gamma=-\Delta^{-1}(\tilde{\gamma}_{t\theta\phi}+F_{\theta\phi}\tilde{\gamma}_{t}\Gamma_{11}).
\end{equation}
Here
$\tilde{\gamma}_i=E_i^A\Gamma_A=\partial_{i}X^{M}E_{M}^{A}\Gamma_A$
are gamma matrices pulled back on the worldvolume and $\Gamma_A$
are the gamma matrices in the tangent space of the background.
There exists a minus sign in (\ref{gamma}) since we consider the
case of an anti-brane. $\Delta$ is defined as
\begin{equation}
\Delta\equiv\sqrt{-det(P[G]_{ab}+{\cal F}_{ab})}.
\end{equation}
For the case at hand, we have
\begin{eqnarray}
\tilde{\gamma}_t&=&\Gamma_{+}+\frac{\mu^2}{18}(R^2+(x^4)^2)\Gamma_{-}, \nonumber \\
\tilde{\gamma}_{\theta}&=&R\Gamma_{\theta}, \nonumber \\
\tilde{\gamma}_{\phi}&=&R\sin\theta\Gamma_{\phi}.
\end{eqnarray}
Using the above relations, the kappa projection can be written as
\begin{equation}
\Gamma=\Delta^{-1}\sin\theta(\Gamma^{-}+\frac{\mu^2}{18}(R^2+(x^4)^2)\Gamma^{+})(R^2\Gamma^{\theta\phi}+q\Gamma^{11}).
\end{equation}
A brane embedding preserves some fractions of the supersymmetry of
the background if the Killing spinor $\eta$ of the background is
consistent with
\begin{equation}
\Gamma\eta=\eta,
\end{equation}
when $\eta$ is restricted on the worldvolume of the brane. The
Killing spinor of the background and the definition of the gamma
matrices are given in the appendix. First we use the kappa
symmetry projection $\Gamma$ to act on the kinematical spinor
$\eta_1$, and we have
\begin{equation}
\Gamma\eta_1=-\sqrt{2}\Delta^{-1}\sin\theta{1 \choose 0}\otimes
(R^2\gamma^{\theta\phi}+q\gamma^9)\tilde{\epsilon}.
\end{equation}
Since it has a different chirality from $\eta_1$, the $16$
kinematical supersymmetries are completely broken. Now we consider
the dynamical Killing spinor. The projection gives
\begin{eqnarray}
-\gamma^{R\theta\phi}(R\gamma^R+x^4\gamma^4)(R^2\gamma^{\theta\phi}-q\gamma^9)\epsilon=\Omega\epsilon, \\
-\gamma^{R\theta\phi}(R\gamma^R-x^4\gamma^4)(R^2\gamma^{\theta\phi}+q\gamma^9)\epsilon=\Omega\epsilon,
\end{eqnarray}
which in turn imply the constraints
\begin{eqnarray}
(R^3+qx^4\gamma^{5678})\epsilon&=&\Omega\epsilon,\nonumber\\
(Rx^4-q\gamma^{5678})\epsilon&=&0,
\end{eqnarray}
where $\Omega=\sqrt{(R^2+(x^4)^2)(R^4+q^2)}$. These constraints
are consistent with $(\gamma^{5678}+1)\epsilon=0$ when evaluated
on the brane worldvolume. So our spherical configuration preserves
all $8$ dynamical supersymmetries of the background. Thus it is a
1/4 BPS state. The moving of giant graviton in one direction and
still preserving supersymmetry remind us of the phenomenon that a
lightcone brane on pp-wave can move along one direction by
boosting or adding fluxes on the worldvolume \cite{braneppwave12,
taylor}. But the mechanism seems different. We notice that the RR
1-form potential can couple to D0-branes in the form $qx^4$ like
the potential of charges in constant electric field. So D0-branes
in this background can feel a force along $x^4$ and thus can move
away from origin until the effects of the gravity increase and
balance it.

Now we are ready to see the spherical D2-brane as the giant
graviton from M-theory point of view. As the extra dimension $x^9$
decompactified in the strong coupling limit, the spherical
D2-brane can be regarded as a spherical membrane in M-theory
pp-wave background. If we further make a coordinate
transformation, the background metric is expressed explicitly in
the form of maximally supersymmetric M-theory pp-wave
(\ref{mppwave1}), and the static membrane becomes rotating in the
plane $X^4-X^9$ with radius $x^4$
\begin{equation}
X^4=x^4\cos(\frac{\mu}{6}x^{+}), \quad
X^9=x^4\cos(\frac{\mu}{6}x^{+}).
\end{equation}
Since the maximally supersymmetric M-theory pp-wave is the Penrose
limit of $AdS_4\times S^7$ background, this rotating spherical
membrane is identified as the 'dual' giant graviton blown up in
$AdS_4$ and circling in $S^7$. From the action of such a rotating
giant graviton on M-theory pp-wave, we can easily get the relation
$R(x^4)^2=2L$, where $L$ is the angular momentum. We first notice
that the radius of the giant graviton is proportional to the
angular momentum as usual. But if the momentum $p=L/x^4$ of the
giant graviton is used, we immediately have $Rx^4\sim p$, which
can be seen as the M-theory counterpart of the relation
(\ref{ryconstraint}) we found in type IIA theory.

Since the giant graviton on IIA pp-wave has no angular momentum
and zero energy, it is not apparent to observe stringy exclusion
principle as shown in \cite{susskind}. However with the essential
relation (\ref{ryconstraint}), we can argue that the existence of
such a giant graviton is actually a manifestation of the spacetime
uncertainty relation in M-theory. From M-theory theory point of
view, the graviton has an energy
\begin{equation}
E\sim {N\over R_c},
\end{equation}
where $R_c$ is the radius of the compact direction $x^9$ and N is
the number of D0-branes attached to it. If we consider the giant
graviton quantum mechanically and use Heisenberg uncertainty
relation, the uncertainty of time is
\begin{equation}
\Delta t\sim {R_c\over N}.
\end{equation}
And from (\ref{ryconstraint}), we know
\begin{equation}
\Delta x^i \Delta x^4\sim l_s^2 N,
\end{equation}
where $i=1,2,3$ and $\Delta x^i\sim R$ is assumed. Thus we deduce
the relation
\begin{equation}
\Delta x^i \Delta x^4 \Delta t\sim R_c l_s^2 \sim l_p^3
\end{equation}
This is just the spacetime uncertainty relation in M-theory
\cite{miaoli,uncertainty}. So from such a connection, we can say
our result about the giant graviton on IIA pp-wave is nontrivial.
The relation (\ref{ryconstraint}) in some sense reflects the
nature of the spacetime geometry.

\section{Microscopic description of giant gravitons}

\indent

Since the spherical D2-brane in the previous section has a
worldvolume magnetic flux which induces a D0-brane coupling and
the background has a 4-form RR field strength, we can consider
this D2-brane to be blown-up from N D0-branes as can be seen from
the Myers effect \cite{myers1}. So we investigate a system of N
D0-branes in type IIA pp-wave background which we expect to give a
microscopic description of the giant graviton found in the
previous section.

As usual the transverse scaler $\phi^i\equiv(2\pi\alpha')^{-1}X^i$
, where $X^i$ represents a matrix valued coordinate, and we only
consider $i=1,...,4$ in the following since other scalars will not
appear in the Wess-Zumino terms. The non-abelian action of N
D0-branes for static configurations on type IIA pp-wave is written
as
\begin{equation}\label{nonabelianaction}
S=-T_0\int dt STr\sqrt{-G_{00}det(Q^i_j)}+\mu_0\int dt
\frac{1}{3}\lambda\mu STr(\phi^4)+\mu_0\int dt
\frac{i}{3}\lambda^2\mu STr(\phi^a\phi^b\phi^c)\epsilon_{abc}.
\end{equation}
Here $Q^i_j=\delta^i_j+i\lambda[\phi^i,\phi^k]$, $
G_{00}=-\frac{\mu^2}{9}(\phi^i)^2$, $\lambda\equiv2\pi\alpha'$ and
$a,b,c=1,...,3$. The symbol $STr$ as usual means the trace is
averaged over all possible orderings of the terms
$[\phi^i,\phi^j]$ and $\phi^k$ appearing inside the trace. In the
following we should consider anti-D0-branes by choosing the RR
charge $\mu_0=-T_0$ to adapt to the anti-D2-brane considered in
the previous section. Written more explicitly, the action
(\ref{nonabelianaction}) becomes
\begin{equation}\label{nonabelian}
S=-\frac{\mu}{3}T_0\lambda\int dt
STr\bigg\{\sqrt{((\phi^4)^2+(\phi^a)^2)(1-\frac{\lambda^2}{2}[\phi^{i},\phi^{j}]^2)}
+\phi^4+i\lambda\phi^a\phi^b\phi^c\epsilon_{abc}\bigg\},
\end{equation}
Since we are interested in the profile of giant graviton blown-up
into the transverse space spanned by $\phi^a$, the scaler $\phi^4$
should commute with $\phi^a$. For $\phi^a$ being an irreducible
representation of some algebra, we can choose
$\phi^4=x^4\identity$. Moreover we assume $|x^4|$ large enough
compared to the size of giant graviton, i.e., $(\phi^a)^2\ll
(x^4)^2$. The validity of the action requires that the commutators
$\lambda[\phi^a,\phi^b]\ll 1$. Expand the square root out in the
action and drop out higher order terms of $\lambda$, the action
can be written as
\begin{eqnarray}
V&\simeq&\frac{\mu}{3}T_0\lambda
STr(-\frac{1}{2x^4}(\phi^a)^2+\frac{\lambda^2}{4}x^4[\phi^a,\phi^b]^2+i\lambda\phi^a\phi^b\phi^c\epsilon_{abc}) \nonumber\\
&=&-\frac{1}{2x^4}\frac{\mu}{3}T_0\lambda
STr(\phi^a-\frac{i}{2}\lambda x^4[\phi^b,\phi^c]\epsilon_{abc})^2.
\end{eqnarray}
Here we have assumed $x^4$ to be negative. So we can see the
energy equal to zero only if
\begin{equation}\label{fuzzysphere}
[\phi^a,\phi^b]=-\frac{i}{\lambda x^4}\epsilon^{abc}\phi^c.
\end{equation}
This shows that D0-branes expand into a fuzzy sphere. But
interestingly transverse coordinate $x^4$ enters into the
commutator as a noncommutative parameter. If we define
$\phi^a=-\frac{J^a}{\lambda x^4}$, then we can see $J^a$ are the
generators of the $SU(2)$ algebra $[J^a,J^b]=i\epsilon^{abc}J^c$.
The matrix $\phi^a$ satisfy
\begin{equation}
(\phi^1)^2+(\phi^2)^2+(\phi^3)^2=\frac{1}{\lambda^2
(x^4)^2}C_2(N)\identity=\frac{N^2}{4}\frac{1}{\lambda^2
(x^4)^2}(1-\frac{1}{N^2})\identity,
\end{equation}
where $C_2(N)=(N^2-1)/4$ is the quadratic Casimir of $SU(2)$ in
N-dimensional representation. This shows that the radius of the
fuzzy sphere should satisfy
\begin{equation}\label{fuzzyrelation}
Rx^4=-\lambda^{-1}N/2.
\end{equation}
Turning to original coordinates $X^i=\lambda\phi^i$ gives
$Rx^4=-\lambda N/2$. Then if we remember that quantization
condition of the magnetic flux requires $N=2\lambda^{-1}q$, we
again arrive at the result $Rx^4=-q$ first exhibited in the
abelian action of D2-brane. For a general solution, $\phi^a$ and
$\phi^4$ can belong to any $N$ dimensional representation of
$SU(2)$ labelled by a partition $\{N_1,...,N_k\}$. It corresponds
to a set of fuzzy spheres with radii $R_i$ and $x^4_i$ satisfying
$R_i x^4_i=-\lambda^{-1}N_i/2$.

In the above analysis we have neglected the high order terms of
$\lambda$ in the nonabelian action. To check the validity of the
solution in the full action, we substitute the fuzzy algebra
\begin{equation}
[\phi^a,\phi^b]=\frac{2R}{N}i\epsilon^{abc}\phi^c
\end{equation}
into (\ref{nonabelian}) and perform the trace. The reduced action
is
\begin{equation}
S=-\frac{\mu}{3}T_2 \int dt d\theta d\varphi \bigg
\{\sqrt{(R^2(1-\frac{1}{N^2})+(x^4)^2)(R^4(1-\frac{1}{N^2})+q^2)}+qx^4-R^3(1-\frac{1}{N^2})\bigg
\}.
\end{equation}
In the large N limit, this action is just the same with
(\ref{abelian}) for spherical D2-brane in type IIA pp-wave
background. So our D0-brane picture do provide a microscopic
description of giant graviton. If we further calculate the energy
of the fuzzy sphere satisfying (\ref{fuzzyrelation}), we find a
remarkable result that it is exactly zero without large $N$ limit.
So the fuzzy sphere is an exact solution of the non-abelian
Born-Infeld action (\ref{nonabelian}). The fact that the energy
corresponding to N D0-branes is cancelled precisely by the
background field indicates that the giant graviton on IIA pp-wave
is actually the condensation of these D0-branes.

It is well known that matrix string theory purports to be a
nonperturbative formulation of string theory. So our giant
gravitons should have a natural description in the matrix string
formulism. Actually, such giant gravitons are just the fuzzy
spheres found recently in the matrix string theory on pp-wave
\cite{shapere}. Here we have a brief discussion about this
solution. Matrix string theory on IIA pp-wave is a (1+1)
dimensional Yang-Mills theory
\cite{matrixstring1,matrixstring2,matrixstring3}. The bosonic
terms of the Lagrangian density are as follows \cite{shapere}
\begin{equation}
{\cal L}=
Tr\bigg\{\frac{1}{2}{g_s}^2{F_{\tau\sigma}}^2+\frac{1}{2}(D_\tau
X^i)^2-\frac{1}{2}(D_\sigma
X^i)^2-\frac{1}{2}(\frac{M}{3})^2(X^a)^2-\frac{1}{2}(\frac{M}{6})^2(X^{a''})^2-\frac{M}{3}g_s
X^4 F_{\tau\sigma}-i\frac{M}{3g_s}\epsilon_{abc}X^a X^b X^c
\bigg\},
\end{equation}
where $M$ is a constant proportional to $\mu$ and $\alpha'$, $g_s$
is the string coupling constant. The indices have a convention
$i=1\cdots 8$, $a=1,2,3$ and $a''=5\cdots 8$. The vacua of the
theory are fuzzy spheres with a translation in $X^4$. Consider the
one fuzzy sphere solution
\begin{equation}\label{fuzzy}
X^i={Mg_s\over 3}J^i, \quad X^4=x^4\identity.
\end{equation}
We notice that this solution also has an electric flux $N$ which
is defined as
\begin{equation}\label{flux}
{1\over 2}\oint d\sigma trE= N,
\end{equation}
where $E$ is the conjugate variable of the gauge potential
$A_\sigma$, and can be solved as a zero energy solution
\begin{equation}\label{e}
E=-{M\over 3}g_s X^4.
\end{equation}
In matrix string theory, the electric flux corresponds to the
D0-brane charge of the configuration. So nonzero $X^4$ indicates
that the fuzzy sphere carries D0-branes. On other hand, from
(\ref{flux}) and (\ref{e}) we have
\begin{equation}\label{x}
x^4=-\frac{3N}{MN_0g_s},
\end{equation}
where $N_0$ is the dimension of the representation. Since $N$ is
an integer, $x^4$ can not be chosen continuously. In other words,
the fuzzy sphere should locate on discrete positions along $X^4$,
which means the $U(1)$ group corresponding to $X^4$ is broken to a
discrete subgroup. If (\ref{fuzzy}) and (\ref{x}) are combined, we
instantly get back to the crucial relation (\ref{fuzzyrelation}).
This result indicates that matrix string theory also provides a
perfect microscopic description of giant gravitons on IIA pp-wave.

\section{Giant gravitons with electric flux}

\indent

In \cite{chen}, it was found that there can be classical stable
BIon configuration with $S^2$ structure as the F-strings bound to
dielectric brane. But the background considered in that paper is
not real supergravity solution and it is worthy to examine such a
profile in a consistent background. In this section we consider
the bound state of a giant graviton and $n$ F-strings on IIA
pp-wave. This configuration can be analyzed by turning on an
electric flux on the worldvolume of a spherical D2-brane. The
embedding can be chosen as
\begin{equation}
x^{+} = t, \quad x^1 = z, \quad x^2 = R \cos\theta, \quad x^3 = R
\sin\theta, \quad F_{tz}=E.
\end{equation}
The Lagrangian (\ref{DBI}) for this embedding can be written as
\begin{equation}
{\cal
L}=-T_2(\sqrt{\frac{\mu^2}{9}(R^2+z^2)(R'^2+1)R^2-E^2R^2}-\frac{\mu}{2}R^2),
\end{equation}
where $R'\equiv\frac{\partial R}{\partial z}$. We can calculate
the Hamiltonian by performing a Legendre transformation
\begin{equation}
{\cal
H}=T_2(\frac{\mu}{3}\sqrt{(R^2+D^2)(R^2+z^2)(R'^2+1)}-\frac{\mu}{2}R^2),
\end{equation}
where $D=\frac{\partial {\cal L}}{\partial E}$ is the conjugate
variable of the electric field. It is noticed that the above
equation can be derived from the membrane action in eleven
dimensional pp-wave background with winding number $n$ along the
compact direction. And it can be seen from this way that $D=ng_s$
with $g_s$ being the string coupling. Since the equation of motion
is quite complicated, we will look for a supersymmetric solution
with the aid of the BPS equation.

As in Section 2, we first write down the kappa symmetry projection
for this embedding
\begin{equation}
\Gamma=\Delta^{-1}(\tilde{\gamma}_{tz\phi}+E\tilde{\gamma}_{\theta}\Gamma_{11}),
\end{equation}
which can be rewritten using gamma matrices in the background as
\begin{equation}
\Gamma=\Delta^{-1}(-R(\Gamma^{-}+\frac{\mu^2}{18}(R^2+z^2)\Gamma^{+})
(\Gamma^{z\theta}+R'\Gamma^{R\theta})-ER\Gamma^{\theta}\Gamma^{11}).
\end{equation}
First the above kappa projection implies that the embedding breaks
the 16 kinematical supersymmetry of the background. The projection
on the dynamical Killing spinor gives
\begin{eqnarray}
&&(\frac{\mu}{3}R\tilde{\gamma}\gamma^{zR}\gamma'+ER\gamma^{\theta
9})\epsilon=\Delta\epsilon, \\
&&(\frac{\mu}{3}R\gamma'\gamma^{zR}\tilde{\gamma}+ER\gamma^{\theta
9})\epsilon=-\Delta\epsilon.
\end{eqnarray}
Here $\tilde{\gamma}\equiv\gamma^{z}+R'\gamma^{R}$ and
$\gamma'\equiv z\gamma^{z}+R\gamma^{R}$. These two equations can
be simplified to
\begin{equation}\label{susy}
\gamma^4\epsilon=\pm\epsilon,
\end{equation}
if the BPS equation
\begin{equation}
RR'+z=\pm\frac{3}{\mu}E
\end{equation}
is satisfied. So we can see this solution preserves $4$ dynamical
supersymmetries of the background. When $E=0$, we have $RR'+z=0$.
This is just the giant graviton we previously discussed. If we
employ $D$ instead of $E$, we have
\begin{equation}
R'=\frac{(\pm D-z)R}{R^2\pm Dz}.
\end{equation}
If $D$ is sufficiently small, the D2-brane solution still
preserves the spherical structure with a deformation only in the
region of small $R$. When we concentrate on the region $R^2\ll
|Dz|$, the BPS equation can be approximate to
\begin{equation}
R'=\mp\frac{R}{D}.
\end{equation}
We can easily integrate it to give
\begin{equation}
R=R_0\exp(\mp\frac{z}{D}),
\end{equation}
where $R_0$ is the integration constant. The sign in the
exponential of the above equation corresponds to the sign in
(\ref{susy}). So the two solutions preserve different factions of
the supersymmetry of the background. It is easily noticed that
these solutions represent two spikes along $\pm z$ direction,
which can be identified with the BI-string carrying a string
charge
\begin{equation}
Q_s=\oint d\theta D=nT_f,
\end{equation}
where $n$ and $T_f$ is the number and tension of the fundamental
string. This result indicates that the BIon configuration with
spherical structure still exists in pp-wave background. Thus we
can explain our configuration as open strings ending on a giant
graviton, giving a realization of the Polchinski's D-brane
picture.

\section{Conclusions and discussions}

\indent

In this paper we have discussed many aspects of the giant graviton
on IIA pp-wave. From the Born-Infeld action, we find that such a
configuration is quite nontrivial: a giant graviton sitting at
different place in $x^4$ has different size. To be precisely this
constraint can be written as $Rx^4=-\lambda N/2$. We can see that
the product of the two transverse coordinates $R$ and $x^4$ is
quantized. If $N$ is fixed, the size of the giant graviton is
inversely proportional to $x^4$. Thus we regard it as a
complementary relation of giant graviton on pp-wave. With this key
relation our giant graviton is shown to be a remarkable
manifestation of the spacetime uncertainty principle. This result
is consistent with previous results on giant gravitons. We also
investigate non-abelian action of D0-branes to give a microscopic
description of the giant graviton. In this case, the complementary
relation is derived naturally from an unusual fuzzy algebra
(\ref{fuzzysphere}), in which we found the coordinate $x^4$
appears as a deformation parameter. This kind of Myers effect has
not been observed before. Further we find the fuzzy sphere is the
exact solution of the full non-abelian action without performing
large $N$ limit. Thus the giant graviton on IIA pp-wave can be
regarded as the condensation of $N$ D0-branes. In the matrix
string theory on IIA pp-wave, we also find the same kind of giant
gravitons in the form of fuzzy spheres. Since the matrix string
theory is argued to be a nonperturbative definition of string
theory, it should allow us to study various aspects of the giant
graviton on IIA pp-wave. This remains to be a future work. In the
last part of this paper, we show a BIon configuration exists on
giant graviton in type IIA pp-wave background. And this result
generalizes our former result in \cite{chen} to a consistent
background.

In the end we should point out that all we have discussed in this
paper is in the background with positive $\mu$ and the giant
graviton in it is made up of an anti-brane. But the case with
negative $\mu$ is also possible. The discussion is parallel if we
consider a spherical brane with a positive RR charge as the giant
graviton.

\bigskip

\acknowledgments

We would like to thank S. Hu, H. Lu and J. X. Lu for helpful
discussion and communication. The work was partly supported by the
NNSF of China (Grant No.90203003) and by the Foundation of
Education Ministry of China (Grant No.010335025).

\appendix
\section{Type IIA pp-wave and Killing spinors}

\indent

In this appendix, we warm up the derivation of the type IIA
pp-wave from the compactification of the maximally supersymmetric
pp-wave in M-theory and give the Killing spinors of this
background. This part is mainly based on \cite{hs2}. The metric we
start with is the M-theory pp-wave with 32 supercharges
\begin{eqnarray}\label{mppwave1}
&&ds^2=-2dX^{+}dX^{-}-(\frac{\mu^2}{9}\sum_{i=1}^{3}(X^{i})^2+\frac{\mu^2}{36}\sum_{i'=5}^{9}(X^{i'})^2)(dX^{+})^2+
\sum_{I=1}^{9}(dX^{I})^2, \nonumber \\
&&F_{+123}=\mu.
\end{eqnarray}
This pp-wave background can be obtained as the Penrose limit of
$AdS_4\times S^7$ or $AdS_7 \times S^4$. To be more intuitively
this is the local geometry seen by a particle circling with
velocity of light in the sphere part of $AdS_4\times S^7$ or
$AdS_7 \times S^4$. It is easily noticed that there are
$SO(3)\times SO(6)$ rotational symmetry of $X^I$ and translation
symmetry of $X^{\pm}$. Besides there are two nontrivial isometries
corresponding to the rotations of the $X^I$ and $X^{-}$ which can
be combined to give a spatial isometry and along which we can
compactify to obtain a type IIA background. To make the isometry
manifest we perform a coordinate redefinition
\begin{eqnarray}
&&X^{-}=x^{-}-\frac{\mu}{6}x^{4}x^{9},\quad \nonumber \\
&&X^4=x^4\cos(\frac{\mu}{6}x^{+})-x^9\sin(\frac{\mu}{6}x^{+}),\nonumber\\
&&X^9=x^4\sin(\frac{\mu}{6}x^{+})+x^9\cos(\frac{\mu}{6}x^{+})
\end{eqnarray}
with other coordinates unchanged. In the new coordinate the metric
reads
\begin{equation}\label{mppwave2}
ds^2=-2dx^{-}dx^{+}-(\frac{\mu^2}{9}\sum_{i=1}^{4}(x^{i})^2+\frac{\mu^2}{36}\sum_{i'=5}^{8}(x^{i'})^2)(dx^{+})^2+
\sum_{I=1}^{8}(dx^{I})^2+(dx^9+\frac{\mu}{3}x^{4}dx^{+})^2.
\end{equation}
Now $x^9$ is a manifest spatial isometry direction. The standard
dimension reduction along this direction gives the type IIA
background
\begin{eqnarray}
&&ds^2=-2dx^{-}dx^{+}-(\frac{\mu^2}{9}\sum_{i=1}^{4}(x^{i})^2+\frac{\mu^2}{36}\sum_{i'=5}^{8}(x^{i'})^2)(dx^{+})^2+
\sum_{I=1}^{8}(dx^{I})^2, \nonumber\\
&&F_{+123}=\mu, \quad F_{+4}=-\frac{\mu}{3}.
\end{eqnarray}
For this background, the vielbein can be chosen as
\begin{eqnarray}
e^{+}=dx^{+}, \quad e^{-}=dx^{-}+\frac{1}{2}A(x^I)dx^{+}, \quad
e^I=dx^I,
\end{eqnarray}
where $A(x^I)\equiv\frac{\mu^2}{9}\sum_{i=1}^{4}(x^{i})^2
+\frac{\mu^2}{36}\sum_{i'=5}^{8}(x^{i'})^2$. The KK gauge field
and the RR 3-form potential can be written as
\begin{equation}
A_{+}=\frac{\mu}{3}x^4, \quad
C_{+ij}=-\frac{\mu}{3}\epsilon_{ijk}x^k.
\end{equation}

This IIA background has 24 supercharges since the toroidal
compactification along a spatial isometry direction inevitably
breaks 8 supercharges \cite{jm}. The 24 supersymmetries of type
IIA pp-wave background are classified into two classes. 16 of them
are non-linearly realized on the string worldsheet and are called
kinematical supersymmetry. The other 8, so called, dynamical
supersymmetry, are linearly realized and time independent. The
kinematical Killing spinor of this type IIA background is
\begin{equation}
\eta_1={0\choose \tilde{\epsilon}},
\end{equation}
where
$\tilde{\epsilon}=e^{-\frac{\mu}{6}\gamma^{123}x^{+}}\tilde{\epsilon}^{+}_{0}
+e^{-\frac{\mu}{3}\gamma^{123}x^{+}}\tilde{\epsilon}^{-}_{0}$ and
$\tilde{\epsilon}^{\pm}_{0}$ are constant spinors satisfying
\begin{equation}
\gamma^{12349}\tilde{\epsilon}^{\pm}_{0}=\pm\tilde{\epsilon}^{\pm}_{0}.
\end{equation}
The dynamical Killing spinor is
\begin{equation}
\eta_2=(1+\frac{\mu}{6}\Gamma^{123}\Gamma^{+}\Gamma^{i}x^{i}){\epsilon
\choose 0}
\end{equation}
Here $\epsilon$ should satisfy $(\gamma^{5678}+1)\epsilon=0$. In
above expressions of Killing spinor, we have chosen the following
representation of the 11-dimensional Gamma matrices
\begin{eqnarray}
\Gamma^0&=&-i\sigma^2\otimes\identity_{16}, \quad
\Gamma^I=\sigma^3\otimes\gamma^I,\quad
\Gamma^9=\sigma^1\otimes\identity_{16}, \quad
\Gamma^{11}=-\sigma^3\otimes\gamma^9, \nonumber \\
\Gamma^{\pm}&=&\frac{1}{\sqrt{2}}(\Gamma^0\pm\Gamma^9),
\end{eqnarray}
where $\sigma$'s are Pauli matrices, and $\identity_{16}$ is the
$16\times 16$ unit matrix. $\gamma^I, {I=1,...,8}$ are real
symmetric gamma matrices satisfying $Spin(8)$ Clifford algebra
$\{\gamma^I,\gamma^J\}=2\delta^{IJ}$. $\Gamma^{11}$ and $\gamma^9$
are defined as
\begin{equation}
\Gamma^{11}=\Gamma^0\cdots\Gamma^9=\Gamma^{-+1\cdots 8}, \quad
\gamma^9=\gamma^1\cdots\gamma^8.
\end{equation}


\end{document}